# Mapping Interdisciplinarity at the Interfaces between the *Science Citation Index* and the *Social Science Citation Index*

Loet Leydesdorff
University of Amsterdam, Amsterdam School of Communications Research (ASCoR),
Kloveniersburgwel 48, 1012 CX Amsterdam, The Netherlands
loet@leydesdorff.net ; http://www.leydesdorff.net

**Abstract**

The two Journal Citation Reports of the *Science Citation Index 2004* and the *Social Science Citation Index 2004* were combined in order to analyze and map journals and specialties at the edges and in the overlap between the two databases. For journals which belong to the overlap (e.g., *Scientometrics*), the merger mainly enriches our insight into the structure which can be obtained from the two databases separately; but in the case of scientific journals which are more marginal in either database, the combination can provide a new perspective on the position and function of these journals (e.g., *Environment and Planning B—Planning and Design*). The combined database additionally enables us to map citation environments in terms of the various specialties comprehensively. Using the vector-space model, visualizations are provided for specialties that are parts of the overlap (information science, science & technology studies). On the basis of the resulting visualizations, "betweenness"—a measure from social network analysis—is suggested as an indicator for measuring the interdisciplinarity of journals.

## 1. Introduction

In another context I was recently asked to contribute to a special issue of *Environment and Planning B—Planning and Design* about information visualization as a potential companion of geo-visualization (Leydesdorff, in preparation). The journal *Environment* *and Planning B* is included in the *Social Science Citation Report,* and using the *Journal Citation Report* 2004 of this index, one can construct a matrix of journals citing the journal within this domain (Table 1).[1]

| | | | | | | | | | | → | citing |
|---|---|---|---|---|---|---|---|---|---|---|---|
| ***Cities*** | 38 | 16 | 7 | 0 | 10 | 0 | 0 | 0 | 2 | 28 | 101 |
| ***Environ Plann A*** | 11 | 228 | 15 | 8 | 32 | 0 | 0 | 13 | 5 | 111 | 423 |
| ***Environ Plann B*** | 2 | 25 | 102 | 34 | 24 | 0 | 0 | 3 | 0 | 22 | 212 |
| ***Int J Geogr Inf Sci*** | 0 | 14 | 13 | 82 | 0 | 0 | 0 | 2 | 0 | 0 | 111 |
| ***J Am Plann Assoc*** | 2 | 6 | 8 | 0 | 60 | 3 | 0 | 4 | 0 | 18 | 101 |
| ***J Archit Plan Res*** | 0 | 0 | 7 | 0 | 6 | 9 | 0 | 0 | 0 | 0 | 22 |
| ***J Urban Plan D-Asce*** | 2 | 7 | 10 | 3 | 15 | 0 | 3 | 0 | 0 | 5 | 45 |
| ***Prof Geogr*** | 3 | 16 | 6 | 13 | 10 | 0 | 0 | 69 | 2 | 14 | 133 |
| ***Prog Plann*** | 13 | 34 | 7 | 21 | 13 | 0 | 0 | 3 | 7 | 33 | 131 |
| ***Urban Stud*** | 23 | 73 | 14 | 2 | 35 | 2 | 0 | 7 | 5 | 250 | 411 |
| ↓ ***cited*** | 94 | 419 | 189 | 163 | 205 | 14 | 3 | 101 | 21 | 481 | 1690 |

**Table 1**: citation matrix of ten journals citing *Environment and Plannning B* within the *Social Science Citation Index 2004.*

On the basis of this matrix, the following map can be generated using the vector-space model for the computation (Salton & McGill, 1983; Ahlgren *et al*., 2003; Leydesdorff, forthcoming):[2]

[1] The ISI aggregates all single citation relations under the category "All others." These missing values are not considered in this analysis.
[2] The visualization program Pajek can be downloaded at http://vlado.fmf.uni-lj.si/pub/networks/pajek/ (Batagelj and Mrvar, 2003; De Nooy *et al*., 2005; Mrvar and Batagelj, *s.d.*).

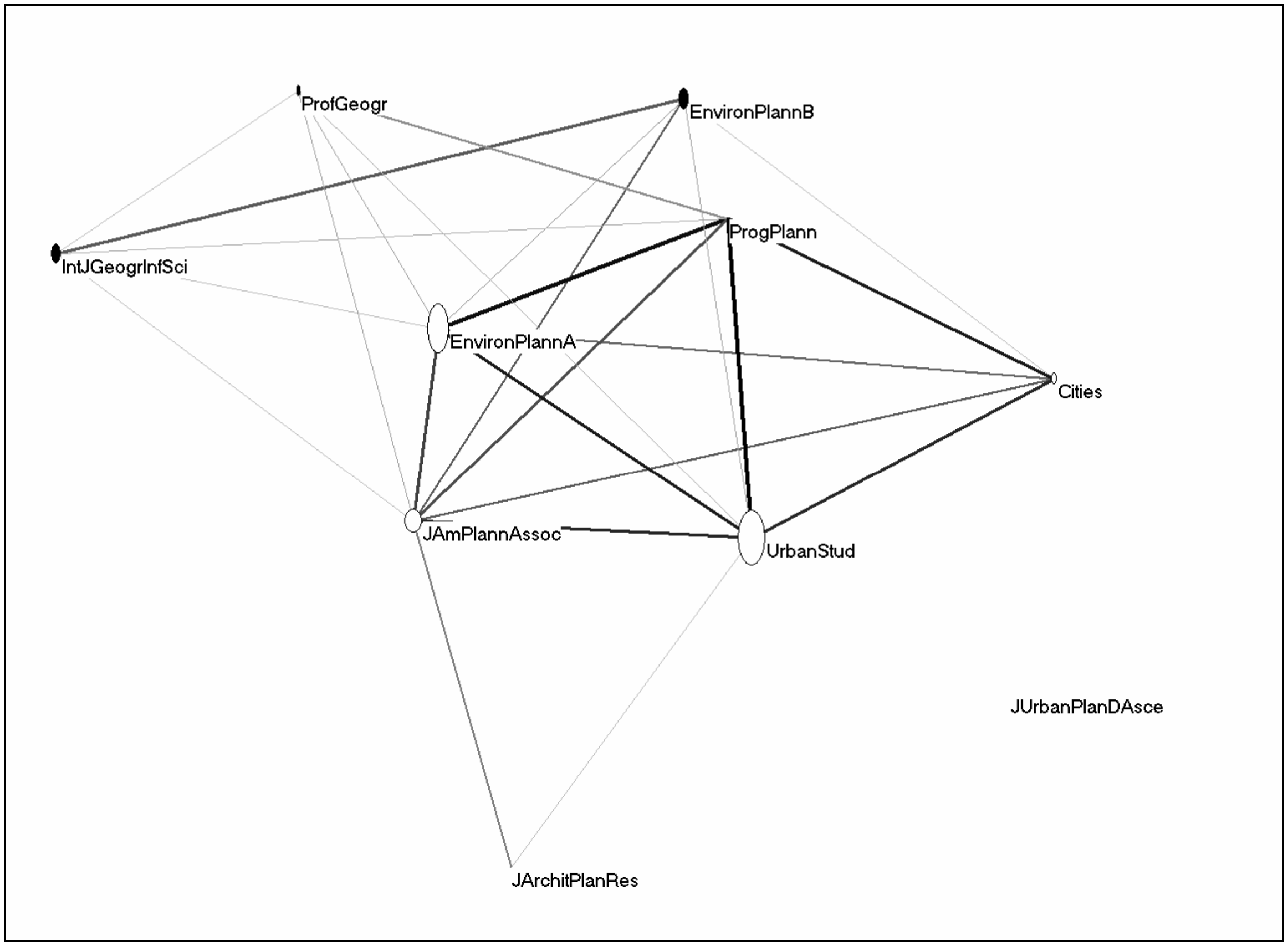

**Figure 1**: Citation impact environment of *Environment and Planning B* in 2004 (cosine ≥ 0.2; see Leydesdorff (forthcoming) for details about the computation and the visualization).

In this figure, the nodes are positioned using a spring-based algorithm which seeks to minimize the 'energy' in the system (Kamada & Kawai, 1980).[3] The width of the lines corresponds to the strength of the relation using the cosine between the vectors (rows in Table 1) as a similarity measure. The vertical size of the node corresponds to the percentage citations that each journal obtains within this local citation environment before correction for within-journal ('self'-)citations, and the horizontal size to this same percentage after correction. Thus, the shapes of the nodes can be considered as indicators of local impacts in the citation environment of *Environment & Planning B.*

[3] This algorithm represents the network as a system of springs with relaxed lengths proportional to the edge length. Nodes are iteratively repositioned to minimize the overall 'energy' of the spring system using a steepest descent procedure. The procedure is analogous to some forms of non-metric multi-dimensional scaling. One disadvantage of this model is that unconnected nodes may remain randomly positioned across the visualization. Unconnected nodes are therefore not included in the visualizations below.

Using the valued core partition algorithm as an option available in Pajek,[4] the position of *Environment and Planning B* in the figure is made visible as more peripheral than the core set of five journals in the area of planning. Together with the *International Journal of Geographical Information* and the *Professional Geographer, Environment and Planning B* forms a subset at the margin of the specialty. (The division is indicated in this figure by using black versus white vertices.) However, this subset could precisely be of interest given our research question of the possible relations between geographical information and information visualization.

The citation environment of journals about visualization techniques includes journals from the computer sciences which are not covered by the *Social Science Citation Index*, but by the *Science Citation Index.* The *Social Science Citation Index 2004* covers 1,712 journals; the *Science Citation Index 2004* includes 5,968 journals. Some journals in information science and the study of science and technology (STS) are included in both databases, but in other cases the divide at the interface seems arbitrary.

For example, *Research Policy,* which can be considered a leading journal in technology studies (Leydesdorff & Van den Besselaar, 1997), is covered by the *Social Science Citation Index,* while *Technology Analysis & Strategic Management* which belongs substantively to the same specialty, is included in the *Science Citation Index.* While journals with "information" in their titles are usually included in both indices, journals in library and information sciences (LIS) that do not have the term "information" in their title are contained only in the *Social Science Citation Index.*

---

[4] This algorithm computes a generalized *k*-core: instead of counting the number of lines, the value of the lines is summed. In accordance with the methodology used in this study, the threshold was set at cosine ≥ 0.2 and the aggregation was stepwise with steps equal to 0.1.

Given the above results, I wondered whether something similar might be the case for *Environment and Planning B*, and hence decided to merge the two databases. The resulting map (Figure 2) was illuminating about the central position of the journal under study. The citation pattern of this journal provides an articulation point between a strong graph of environmental-planning journals and a graph of computer-science journals.[5] The *k*-core algorithm, however, attributes the relatively marginal cluster in the previous figure as belonging to the computer-science component in this configuration.

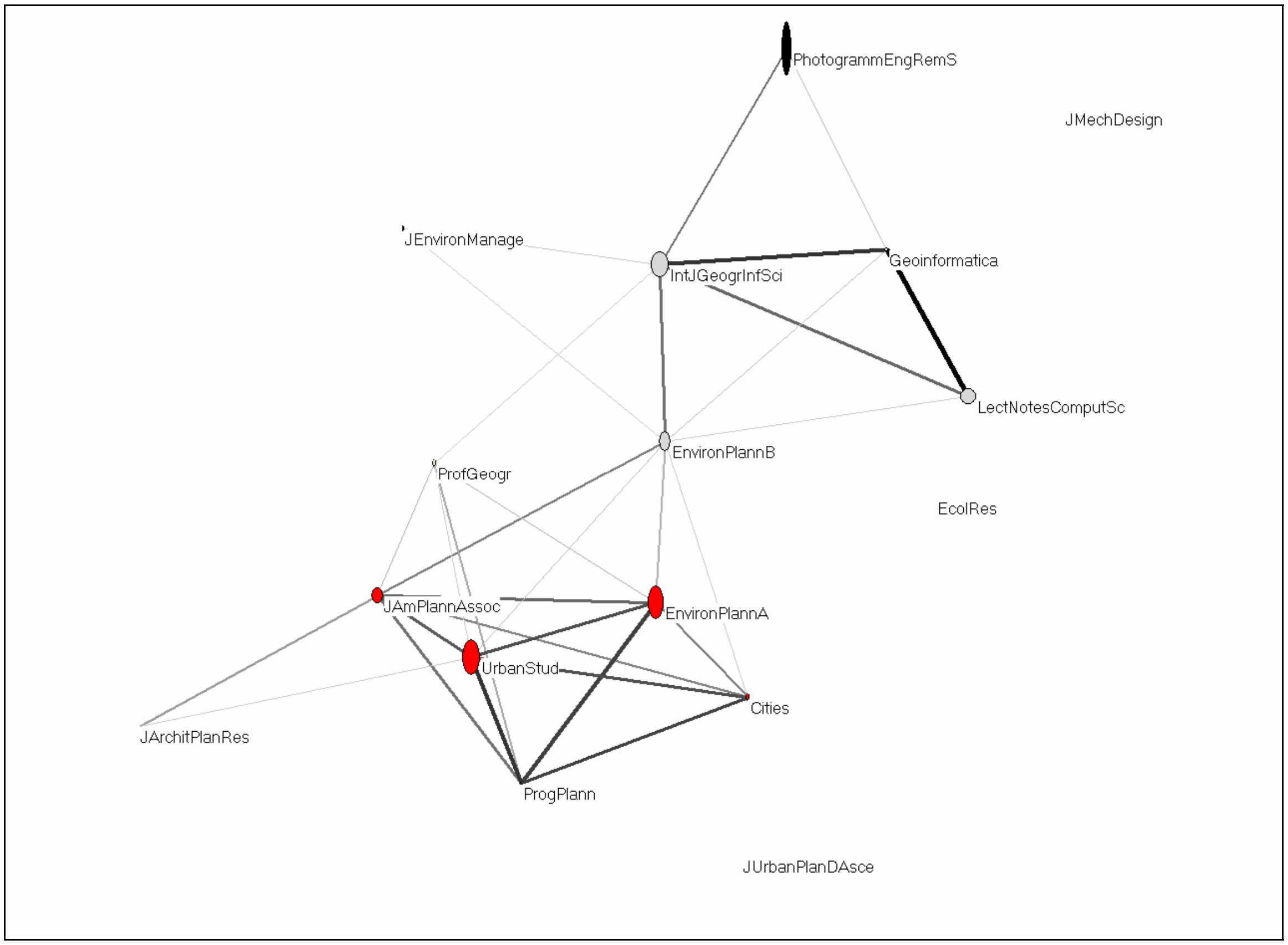

**Figure 2** Local citation impacts of journals in the citation environment of *Environment and Planning B* in 2004. (*Social Science Citation Index* and *Science Citation Index* combined; cosine $\geq$ 0.2.) [6]

[5] "Articulation points" or "cut-points" are defined in graph theory as vertices that are shared between two bi-connected components (Scott, 1991).

[6] The size of the node for *Lecture Notes in Computer Sciences* has been reset to one percent because this journal has a citation rate of 32,739 citations (among which 18,005 within-journal citations), but these citations are not provided in this environment.

This result raises the question of how a mapping combining the two databases would work for information-science journals like *JASIST* and *Scientometrics*. In the remainder of this paper, I first show—using *JASIST* as the seed journal—that the results of the analysis in the two combined databases do not add to our understanding of the position of the information sciences at the interface when compared with the results of the two separate analyses. This raises issues about the different definitions of "interdisciplinarity" in the case of individual journals (like *Environment and Planning B*) or groups of journals representing specialties which function at the interface.

Thereafter, I shall exploit an additional advantage of the combined database. The combined set enables us to take a bird-eye's view of the larger environment of a field, including all citing journals, that is, a cited journal's complete and potentially interdisciplinary citation impact environment. The results will show the different positions of leading journals in information science and STS between the two databases, and accordingly their potentially different functions at relevant interfaces (Leydesdorff & Van den Besselaar, 1997).

## 2. Methods and materials

### *2.1 Materials*

The data was harvested from the *Journal Citation Reports* of the *Science Citation Index* and the *Social Science Citation Index 2004.* The descriptive statistics of the two databases and the effects of their combination are provided in Table 2. The two databases contain 5,968 and 1,712 journals, respectively, or a total of 7,680 journals. However, the overlap is (7,680 – 7,379 =) 301 journals. Only seven journals in this overlap are not processed by the ISI.[7]

[7] Not all source journals are processed actively, that is, in the citing dimension. In the *Science Citation Index 2004,* 192 journals were not actively processed in this dimension. However, 24 of them had no citations at all. The other 168 journals are only registered when cited by other journals. The corresponding

| | *SCI 2004* | *SoSCI 2004* | *Combined* | *Effect* |
|---|---|---|---|---|
| Number of source journals | 5968 | 1712 | 7379 | +28.6% |
| unique journal-journal relations | 1,038,268 | 96,207 | 1,195,158 | +15.1% |
| sum of journal-journal relations | 18,943,827 | 966,619 | 20,326,793 | +7.3% |
| *average cell value* | *18.25* | *10.05* | *17.01* | *-6.7%* |
| total ‘citing’ | 25,798,965 | 2,909,219 | 27,961,981 | +8.4% |
| total ‘cited’ | 20,909,401 | 1,453,397 | 21,810,032 | +4.3% |
| within-journal citations’ | 2,016,500 | 137,269 | 2,107,885 | +4.5% |

**Table 2**: Descriptive statistics of the JCRs of the *Science Citation Index,* the *Social Science Citation Index*, and the two databases combined.

The overal impression from the figures in Table 2 is that the extension of the *Science Citation Index* by the *Social Science Citation Index* with 28.6% (= 1,712/5,968) more source journals has an effect on the citation characteristics of the set of less than ten percent. Thus, the database is changed, but one would expect the structures of the *Science Citation Index* to persist in the combined set.

### *2.2 Methods*

A citation index contains all information for the construction of a huge matrix in which the cited journals provide information for the row vectors and citing journals for the column vectors (or *vice versa*); the cell values are equal to the number of unique citation relations at the article level. The matrix is asymmetrical, and the main diagonal— numbers of inactive journals were 40 in the *Social Science Citation Index,* and 199 in the combined database.

representing “within-journal” citations—provides an outlier in the otherwise skewed distributions.[8]

As the similarity measure between the distributions for the various journals included in a citation environment, I use the cosine between the two vectors or, in other words, normalization to the geometrical mean. Unlike the Pearson correlation coefficient, the cosine does not presume normality of the distribution (Ahlgren *et al*., 2003). A further advantage of this measure is its further development into the so-called vector-space model for the visualization (Salton & McGill, 1983).

A citation environment is defined as all journals which cite or are cited by a specific journal above a given threshold. The value of this threshold can be varied, but the default value is set at one percent of the total references or citations in the citing and cited dimensions of the matrix (He & Pao, 1986; Leydesdorff & Cozzens, 1993). When the threshold is set equal to zero, one is able to map the complete citation context of a journal.

The cosine matrices (with the default value for the threshold) of all journals included in the *Science Citation Index* and the *Social Science Citation Index* combined are brought online at http://www.leydesdorff.net/jcr04/cited and http://www.leydesdorff.net/jcr04/citing, respectively, in a format which allows a user to generate maps as in Figure 2 above by using Pajek or to export this data for statistical analysis from Pajek into UCINET and SPSS.[9]

## 3. Results

### *3.1 Information science & technology*

Figures 3, 4, and 5 show the citation impact environments of the *Journal of the American Society of Information Science and Technology (JASIST)* in the *Social Science Citation*

---

[8] Within-journal citation traffic accounts for about 10% of the total citation traffic (Leydesdorff, forthcoming).
[9] Pajek is freely available for non-commercial usage at http://vlado.fmf.uni-lj.si/pub/networks/pajek/ .

*Index,* the *Science Citation Index*, and the two indices combined. In the network from the *Social Science Citation Index*, *JASIST* is itself the leading journal in an otherwise strong graph of journals in the library and information sciences (Figure 3). The somewhat different position of *Scientometrics* in this environment is indicated from the perspective *JASIST* as the seed journal, but cannot further be clarified without focusing on this journal's own citation environment. I shall turn to the citation environment of *Scientometrics* in a later section.

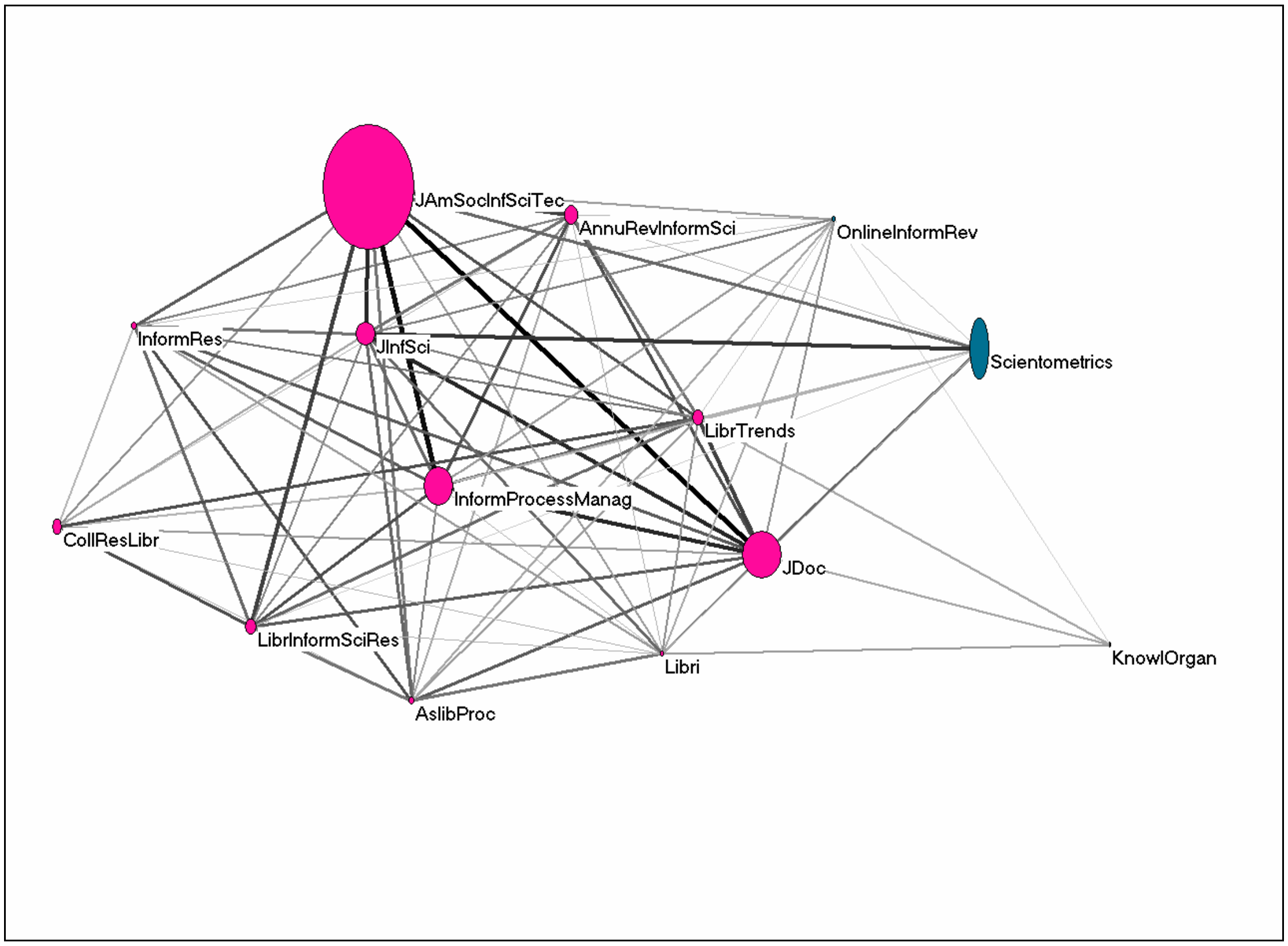

**Figure 3:** Citation impact environment of the *Journal of the American Society of Information Science & Technology* 2004 on the basis of the JCR of *SoSCI.*

In the environment of the *Science Citation Index* (Figure 4), however, two larger journals become visible: the *Proceedings of the National Academy of the U.S.A. (PNAS)* and *Lecture Notes in Computer Science.* The citation relation with the *PNAS* in 2004 is perhaps an artifact of the special issue of this journal on the subject of the visualization of knowledge, to which a number of authors from the bibiometric community contributed (Shriffin & Börner, 2004). The relations with *Lecture Notes in Computer Science* and *Lecture Notes in Artificial Intelligence,* however, manifest a structural part of the graph. *JASIST* and even more *Information Processing and Management* seem to function at the interface between information science and the computer sciences.

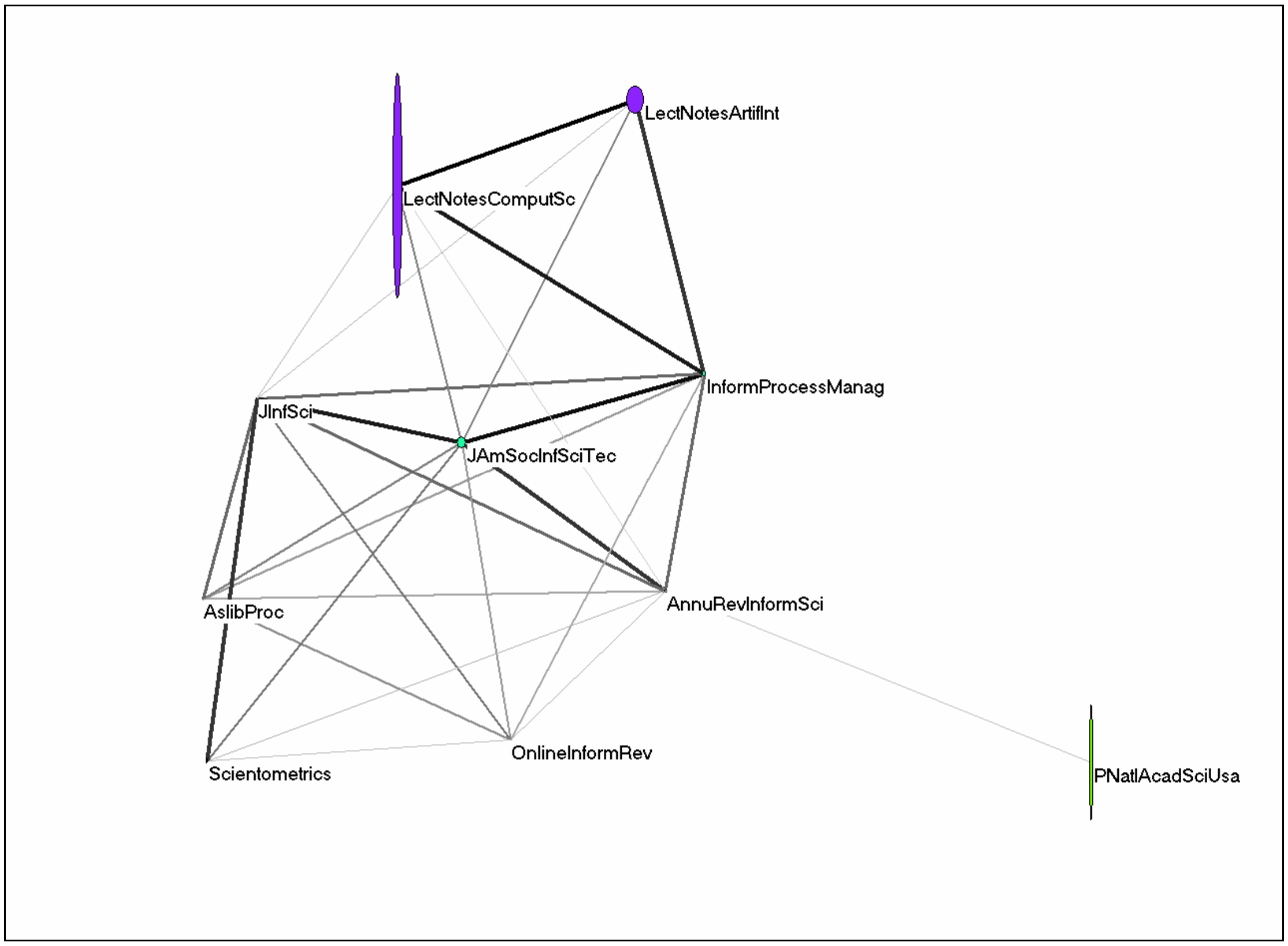

**Figure 4**: Citation impact environment of the *Journal of the American Society of Information Science & Technology* 2004 on the basis of the JCR of *SCI.*

The merger of the two databases enables me to generate additionally the datamatrix for Figure 5. Comparison of Figure 5 with Figure 4 shows that the addition of the *Social Science Citation Index* informs the picture, but structurally the two figures remain similar: the graph of the information-science journals in the lower half of the picture is further informed by adding the library-science journals, but not changed in shape. Figure 3 enabled us to visualize this fine structure among LIS journals in greater detail, but it provides a visualization of only a substructure of Figure 5 and, therefore, Figure 5 does not really add to our understanding of the intellectual organization around and among LIS journals. For the purpose of mapping and visualization, however, this figure is more complete.

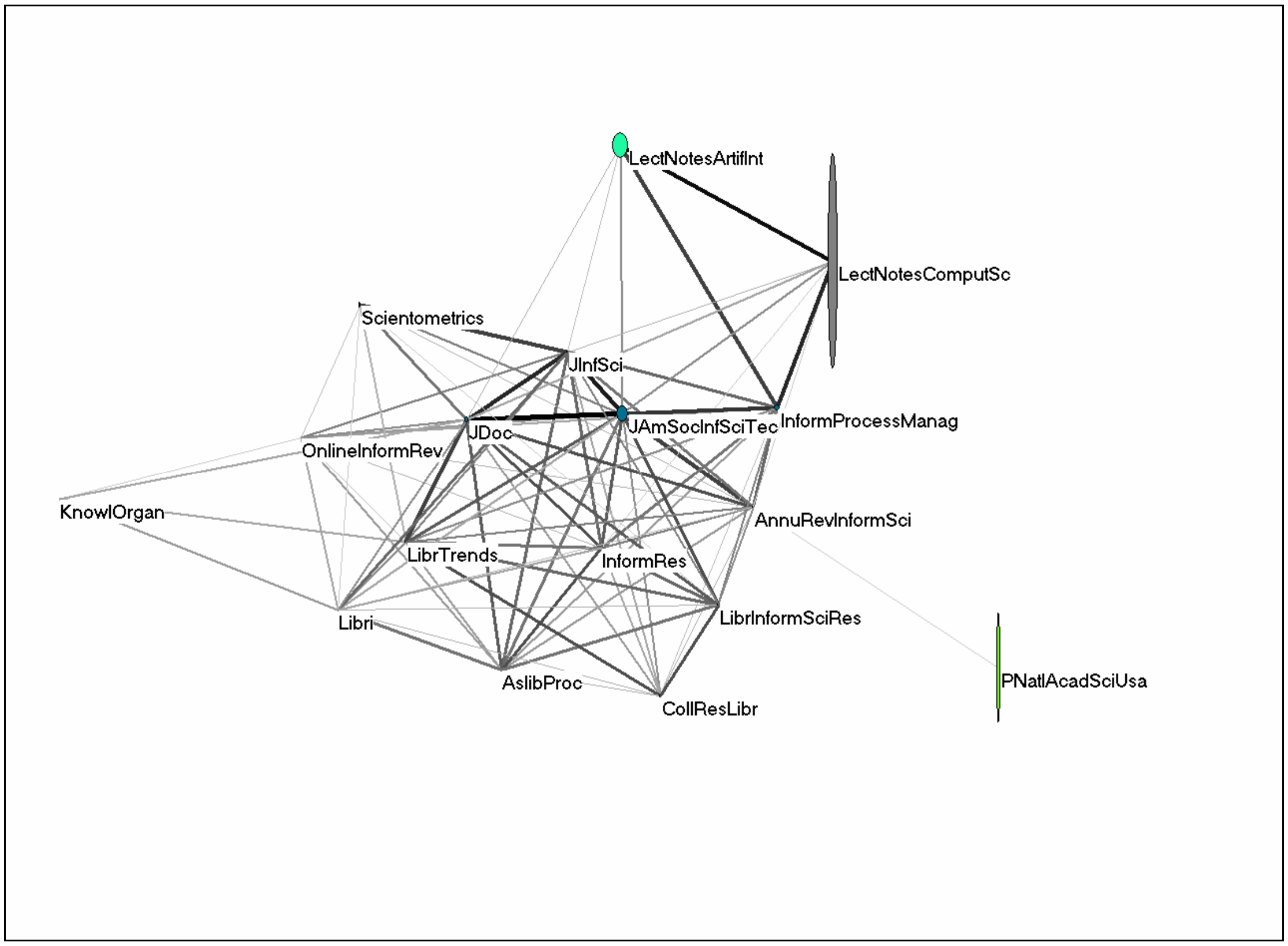

**Figure 5**: Citation impact environment of the *Journal of the American Society of Information Science & Technology* 2004 on the basis of the JCRs of the *SCI* and the *SoSCI* combined.

In other words, while we were able to understand the marginal position of *Environment and Planning B* in the context of the *Social Science Citation Index* because of its specific role as a *journal* at the interface with the computer sciences, this function at the interface is carried by the *set* of information-science journals which form a strong graph in terms of the citations among them. However, it may be interesting to visualize how this set is positioned among other sets of journals in the two databases, and in the context of the two databases combined. As noted above, this can be done by reducing the threshold for inclusion to zero. In the next section, I include all the journals citing the seed journal.

## 3.2 The larger environments

Figure 6 provides the citation impact environment of *JASIST* when all 105 journals citing *JASIST* are included in the analysis. However, seventeen journals are not included in the visualization because these journals do not pass the similarity criterion of a cosine greater than or equal to 0.2 with at least one other journal in the set.[10]

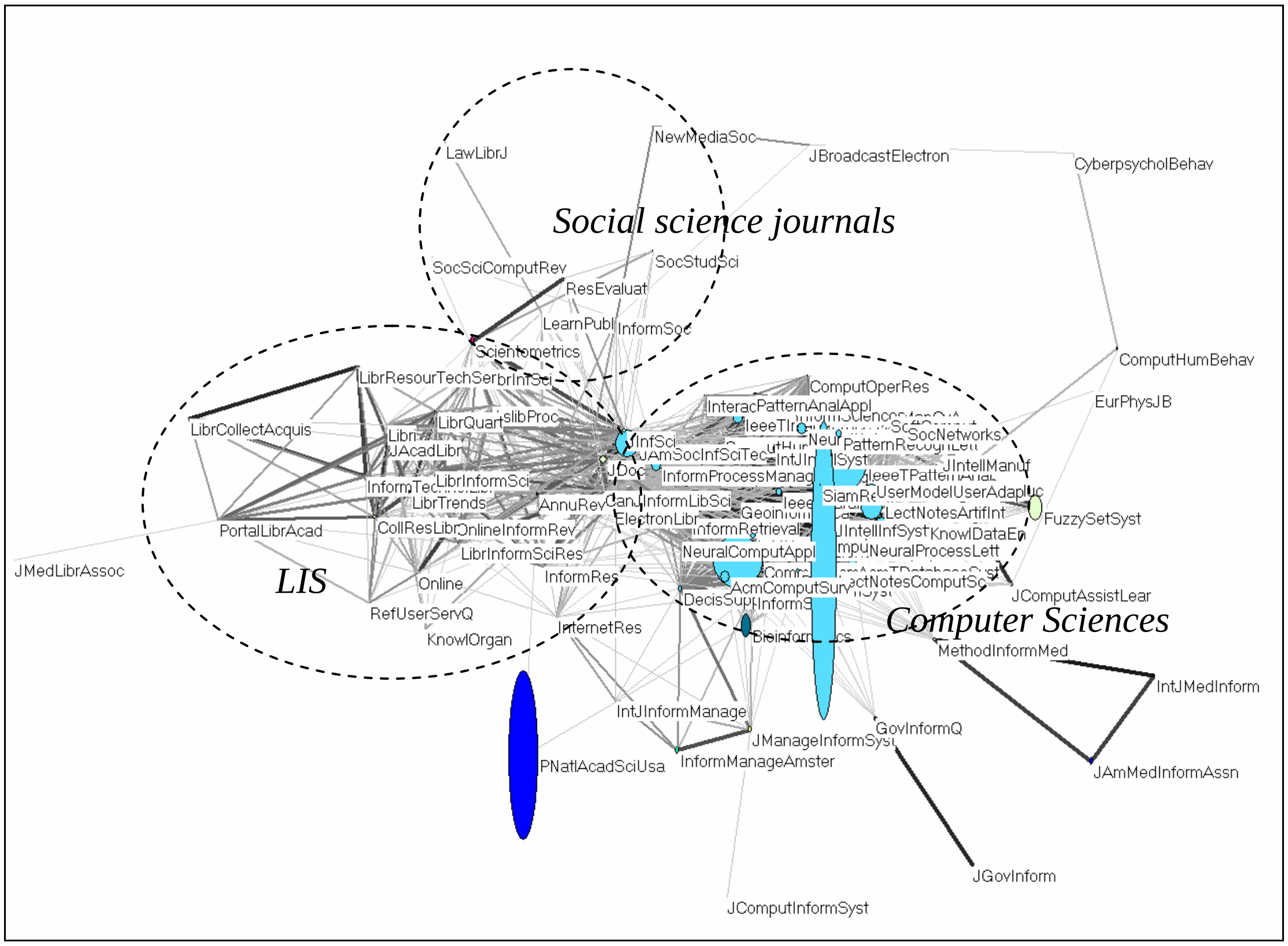

**Figure 6**: Citation Impact Environment of *JASIST* of 88 journals in its local citation environment (cosine ≥ 0.2)

The journals in the library and information sciences form a tightly knit cluster; the graph is stronger on the side of the information sciences than on the side of the library sciences. (This was also visible in Figures 3 and 5 above.) At the information-science end, the graph overlaps with journals in the computer sciences. The *k*-core algorithm of Pajek even indicates *JASIST* and *Information Processing and Management* as belonging to the graph of the computer sciences more than library and information science.

[10] Without a threshold in the similarity criterion, the graph would be completely saturated because the cosine between two vectors with positive values is always larger than zero.

A strong group of approximately one hundred journals can be retrieved by using any major journal in the computer sciences as a seed journal. Since *Lecture Notes in Computer Science* has an exceptionally large citation environment of 585 journals, I used *Lecture Notes in Artificial Intelligence* as the seed journal. This journal was the other computer-science journal visible in the restricted environment of *JASIST* (see Figure 5).

*Lecture Notes in Artificial Intelligence* is cited by 144 journals, of which 132 pass the threshold for the visualization. In Figure 6, various specialist areas are visible at the edges of the large and dense group of computer-science journals, including information-science journals. *JASIST* and *Information Processing and Management* are again integrated in this graph, while the *Annual Review of Information Science* and the *Online Information Review* are positioned at a distance. As noted, a similar picture of this dense graph of approximately one hundred journals in the computer sciences can be generated using almost any of these journals as a seed journal.

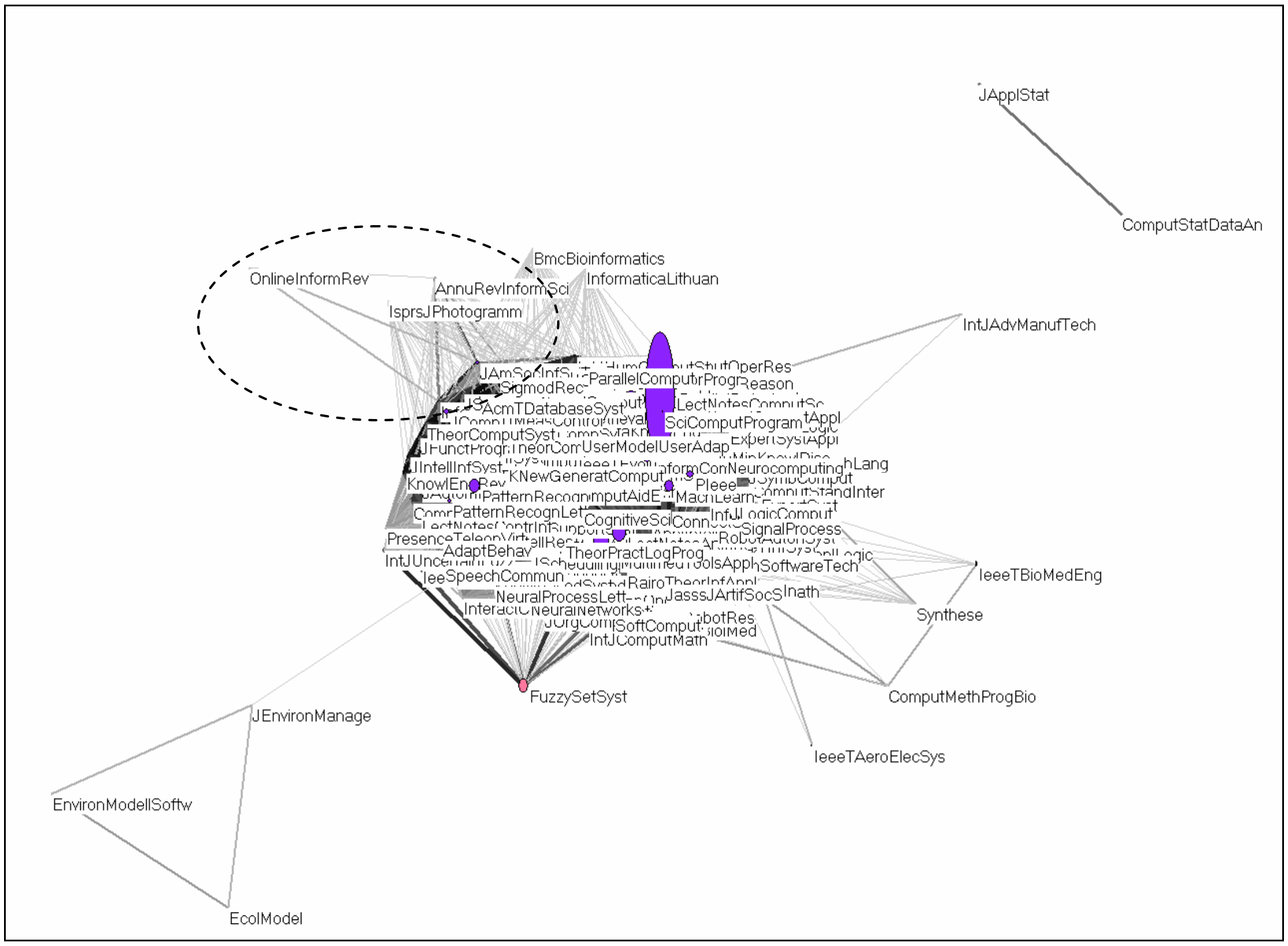

**Figure 7**: Citation Impact Environment of *Lecture Notes in Artificial Intelligence* on 132 journals in its local citation environment (cosine ≥ 0.2).[11]

*Scientometrics* is not part of the last figure, nor of the *LIS*-graph in Figure 6. In the LIS environment, *Scientometrics* functions as an articulation point with a group of social science journals, among them *Social Studies of Science*. However, if we use *Scientometrics* itself as a seed journal (Figure 8), these journals are not drawn into its citation impact environment. *Scientometrics* has a citation impact in the information sciences and in a group of journals oriented towards S&T policy issues and R&D management. Authors in these journals use S&T-indicators in their studies and for this reason provide references to *Scientometrics* and *Research Evaluation*. *JASIST*, however, has citation impact in journals like *Information Society* and *Social Studies of Science*, which also consider the social aspects of information and communication technologies.

[11] Eleven more journals play a role in this citation environment, but are not connected to the graph at the level of cosine ≥ 0.2.

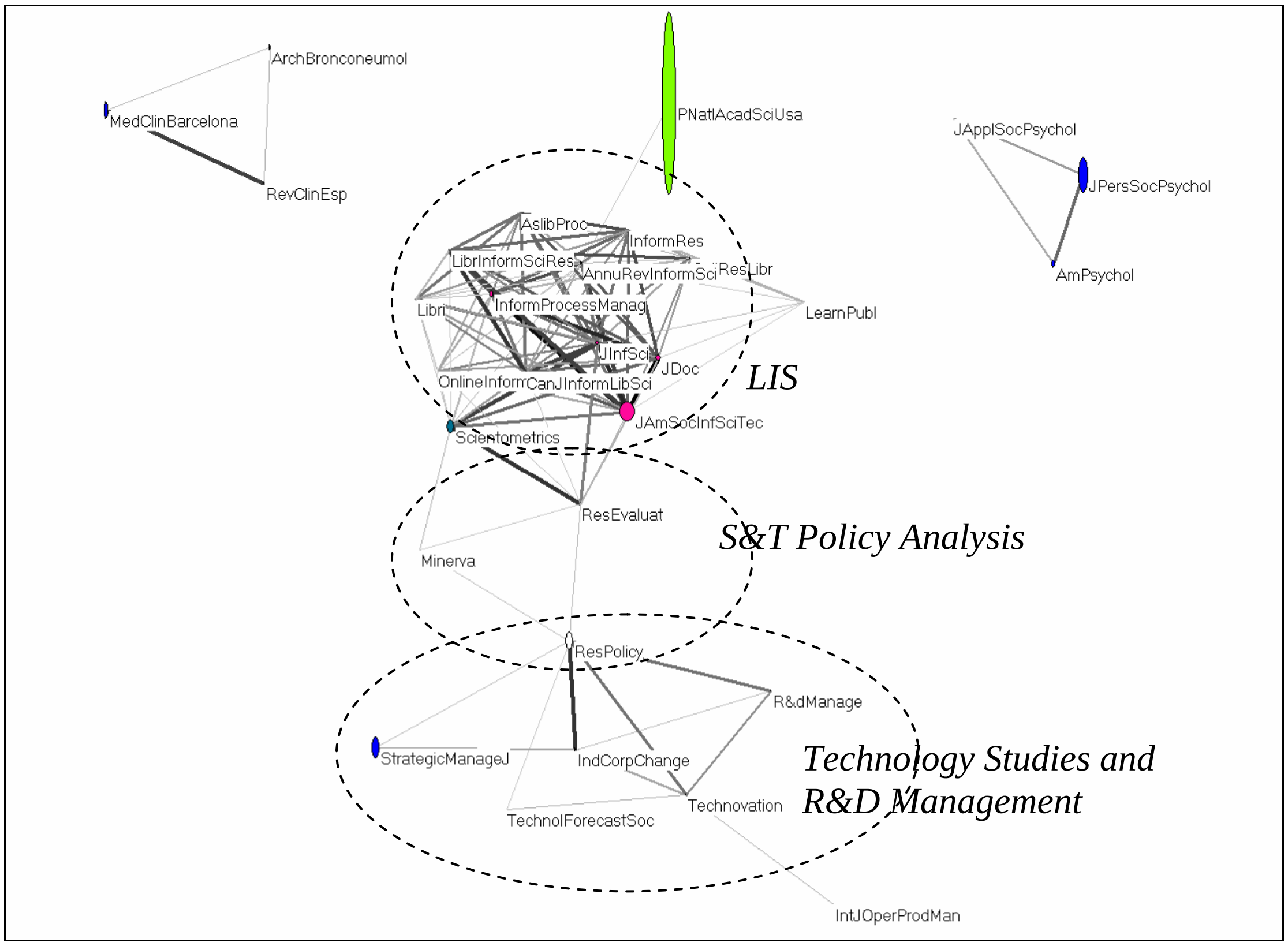

**Figure 8**: Citation Impact Environment of *Scientometrics* on 30 journals in its local citation environment (threshold = 0.1%; cosine ≥ 0.2)

Twenty-four of the fifty-four journals which cite articles in *Scientometrics* more than once do not pass the threshold of the similarity measure for the visualization (cosine ≥ 0.2). This is a relatively large group (44%), including such journals as *Current Science* and the *International Journal of Bifurcation and Chaos.* These journals sometimes accept papers of bibliometricians, or are otherwise (but not systematically) related to the field. Note that the *Strategic Management Journal* was part of the citation impact environment of *Scientometrics* in 2004 although it belongs to the field of the management sciences.

Let us now turn to two central journals in Science & Technology Studies: *Research Policy* and *Social Studies of Science*. The latter can be considered as a leading journal in the sociology of scientific knowledge, while the former is the leading journal at the interface between technology studies and evolutionary economics. Leydesdorff & Van

den Besselaar (1997) noted already that these two journals no longer maintain citation traffic between them although they originated in the 1970s and 1980s from the same community (Leydesdorff, 1989; Van den Besselaar, 2001). *Social Studies of Science* is not included in the complete citation environment of *Research Policy* (zero citations), and the latter journal is cited in the former's citation environment only thirteen times (of a total of being cited 428 times in this environment). Thus, the relations between these journals are marginal and the patterns of their citations are not significantly correlated.[12]

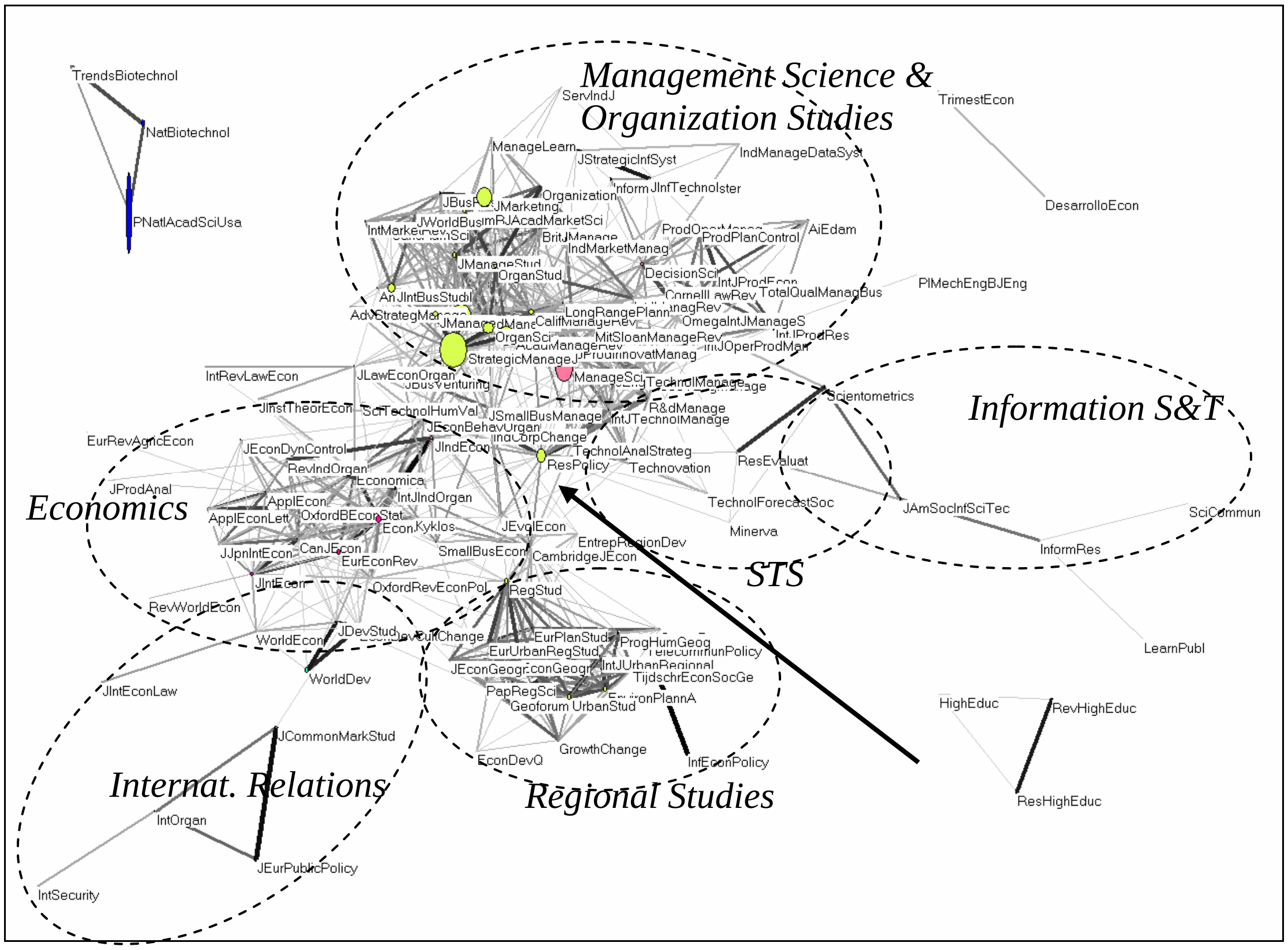

**Figure 9**: Citation Impact Environment of *Research Policy* on 122 journals in its local citation environment (cosine ≥ 0.2)

Articles in *Research Policy* were cited during 2004 in 139 other journals, of which 122 are connected at the level of a cosine larger than or equal to 0.2 (Figure 9). Different fields surround the seed journal which is positioned in the middle and indicated in this

[12] The Pearson correlation coefficient *r* between the citation patterns of the two journals in the citation environment of *Social Studies of Science* is 0.115 (n.s.), and Spearman's ρ = 0.175 (n.s.).

figure with an arrow. Together with certain other journals (like *Industry and Corporate Change* and the *Journal of Evolutionary Economics*), *Research Policy* functions at a crossroads between institutional economics, STS, regional studies, and management studies. Journals with a focus on information science and technology at the one end, and international relations at the other, are more weakly related to the seed journal.

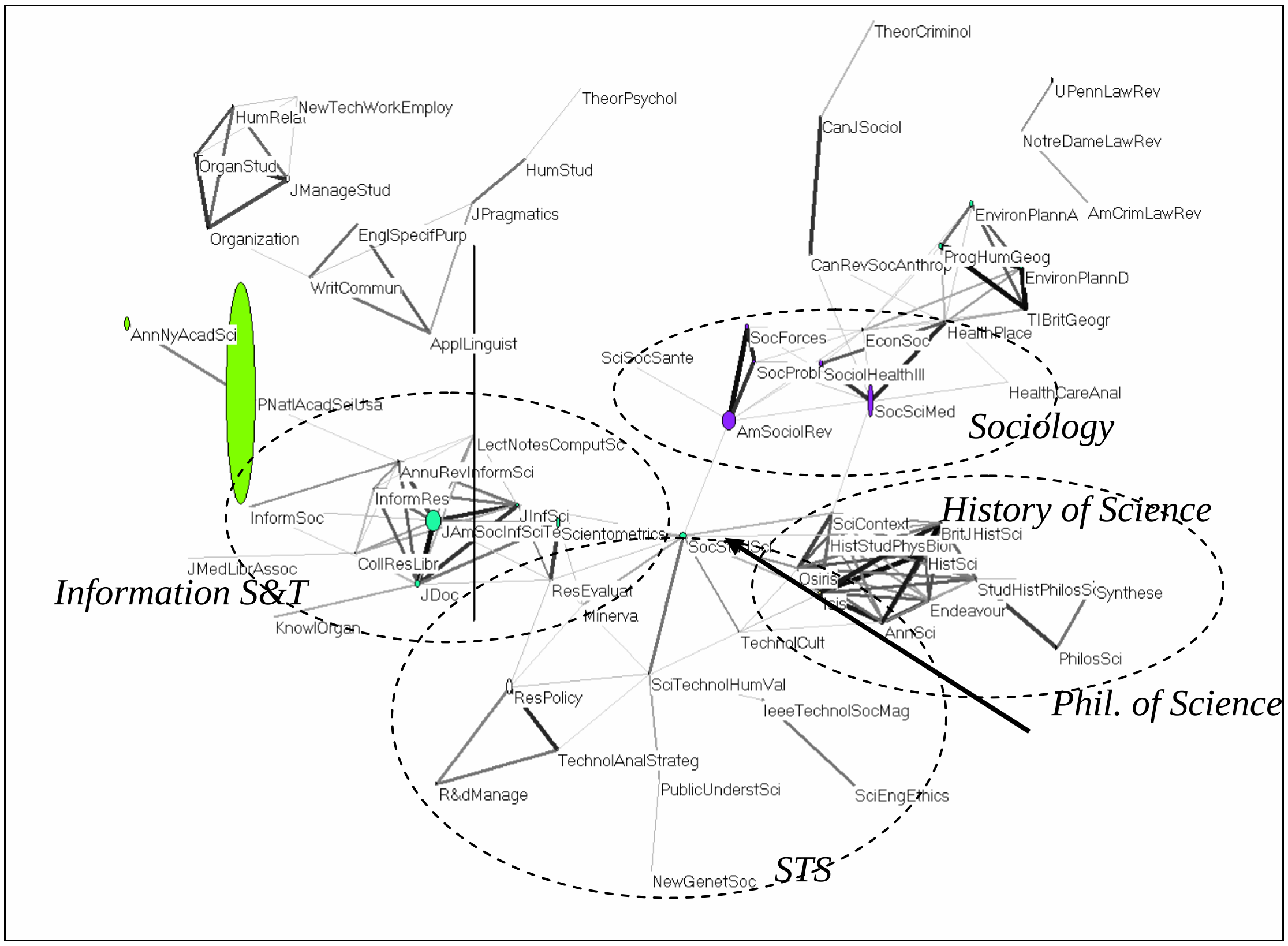

**Figure 10**: Citation Impact Environment of *Social Studies of Science* on 66 journals in its local citation environment (cosine ≥ 0.2)

Similarly—in graph-theoretical terms—*Social Studies of Science* occupies a structural position in the center of Figure 10 between the history and philosophy of science, STS, sociology, and the information sciences. *Research Evaluation* and *Scientometrics* again mediate the latter relation. Of the 81 journals citing *Social Studies of Science* in the two databases, 66 are connected among themselves in terms of their citation patterns above the threshold value of cosine ≥ 0.2.

The star-shaped position of the seed journals among various specialties in the last two analyses (Figures 9 and 10) raises the question of whether a graph-theoretical indicator of interdisciplinarity can be formulated on the basis of this pattern. *JASIST, Lecture Notes in Artificial Intelligence,* and *Scientometrics* did not show this specific pattern in the respresentations of their local citation environments. *Research Policy* and *Social Studies of Science* remain with their citation impact largely within the domain of the *Social Science Citation Index*. *Lecture Notes in Artificial Intelligence* integrates journals which typically belong to the social sciences (e.g., *JASSS—Journal of Artificial Societies and Social Simulation*) into the strong graph of computer-science journals. The combination of the two databases did not affect this cluster structurally.

## 4. Conclusion

The merger of the two Journal Citation Reports of the *Science Citation Index* and *Social Science Citation Index* offers an opportunity to map scientific specialties and disciplines across this interface. The evaluation of a social science journal which is linked to or integrated into the *Science Citation Index* may change dramatically by using this different context. If a journal is rather marginal in one set—as in the case of *Environment and Planning B*—the addition of the other database can provide interesting perspectives on its position in the field and its function in the network. In the case of the information sciences, however, the resulting insight into the position of this *group* of journals was not very different from the results of the two databases separately. The evaluation of STS journals was not much affected because they belong structurally to the domain of the social sciences.

The merger enables us to generate a bird-eye's view of the position of interdisciplinary groups of journals in the two databases because the mapping can be extended to the complete citation environment. The complete citation environments for most specialist journals are of the order of one hundred journals. Local citation impact can be comprehensively visualized in such relatively small environments. I envisage bringing the citation impact environments online in 2005 without setting a citation threshold.

Furthermore, these results suggest that measures from social network analysis like "betweenness" or one of its derivates can be used as a quantifiable indicator of the interdisciplinarity of journals (Freeman, 1978/1979; Wasserman & Faust, 1994; Hanneman & Riddle, 2005). However, whether such an indicator should be applied to the citation matrix itself or to the cosine matrix representing the vector space requires further analysis. It might well be that different centrality measures have to be used for measuring "interdisciplinarity" or "multidisciplinarity" (e.g., *Nature*). Additionally, "interdisciplinarity" needs to be defined differently for the cited and the citing dimensions (Leydesdorff, 1993).

**Acknowledgement**

I am grateful to Caroline S. Wagner for useful comments on a previous draft.